\documentclass{jfm}
\usepackage{graphicx}
\usepackage{amsmath}
\usepackage{epstopdf, epsfig}
\usepackage{float}
\usepackage{amssymb}
\usepackage{color,contour}
\usepackage{bm}

\usepackage[colorlinks,citecolor = blue, linkcolor=red,hyperindex,CJKbookmarks]{hyperref}
\newcommand{\makered}[1]{\textcolor{black}{#1}}

\shorttitle{Lagrangian dynamics and heat transfer}
\shortauthor{S. Liu, L. Jiang, C. Wang, and C. Sun}

\title{Lagrangian dynamics and heat transfer in porous-media convection}

\author{Shuang Liu\aff{1},
	Linfeng Jiang\aff{1},
	Cheng Wang\aff{1},
	Chao Sun\aff{1,2}\thanks{Email address for correspondence: chaosun@tsinghua.edu.cn}}

\affiliation{
	\aff{1}Center for Combustion Energy, Key Laboratory for Thermal Science and Power Engineering of Ministry of Education, Department of Energy and Power Engineering, Key Laboratory of Advanced Reactor Engineering and Safety of Ministry of Education, Tsinghua University, Beijing 100084, China
	\aff{2}Department of Engineering Mechanics, School of Aerospace Engineering, Tsinghua University, Beijing 100084, China
}
\begin{document}

\maketitle

\begin{abstract}
We report a numerical study of Rayleigh--B{\'e}nard convection through random porous media using pore-scale modelling, focusing on the Lagrangian dynamics of fluid particles and heat transfer for varied porosities $\phi$. Due to the interaction between the porous medium and the coherent flow structures, the flow is found to be highly heterogeneous, consisting of convection channels with strong flow strength and stagnant regions with low velocities. The modifications of flow field due to porous structure have a significant influence on the dynamics of fluid particles. Evaluation of the particle displacement along the trajectory reveals the emergence of anomalous transport for long times as $\phi$ is decreased, which is associated with the long-time correlation of Lagrangian velocity of the fluid. As porosity is decreased, the cross-correlation between the vertical velocity and temperature fluctuation is enhanced, which reveals a mechanism to enhance the heat transfer in porous-media convection.
\end{abstract}

\vspace{-15 mm}

\section{Introduction}
Transport and mixing processes in porous-media flows have attracted much attention over the years, owing to their importance in a wide range of natural and industrial settings, such as the contaminant transport in the subsurface, the kinetics of chemical reactions, and the transport in biological systems \citep{manke2007investigation,seymour2004anomalous,cushman2016handbook,gu2019longitudinal,wu2019study}. 
Understanding the dynamics of fluid particles in complex flows is also important from the theoretical perspective, since the features of the fluid flow advecting the particles can be inferred from the particle dynamics \citep{falkovich2001particles,biferale2004multifractal,toschi2009lagrangian,calzavarini2020anisotropic,mathai2020bubble}.

Significant progress has been achieved for the transport and mixing processes in the pressure-driven porous-media flows in the Darcy regime. 
The transport dynamics are governed by the probability density function (p.d.f.) of velocity in the pores, particularly the distribution in the low-velocity range, which plays a critical role in the anomalous or non-Fickian transport behaviour in porous media \citep{berkowitz2006modeling}. A large probability of low velocities can result in persistent anomalous transport.
The anomalous transport behaviours in heterogeneous flow field have been modelled using the continuous-time random-walk approaches, accounting for the impact of broad distributions of advective times in the pores \citep{berkowitz2006modeling,bijeljic2006pore,bijeljic2011signature,bijeljic2013predictions,de2013flow,kang2014pore,lester2014anomalous,holzner2015intermittent,dentz2016continuous,morales2017stochastic,nissan2018inertial,dentz2018mechanisms,puyguiraud2019stochastic,puyguiraud2019upscaling,souzy2020velocity}.
{\color{black}\cite{dentz2018mechanisms} identified and quantified the role of advection and molecular diffusion on the preasymptotic non-Fickian transport, and found that the non-Fickian transport features can persist on the scale of representative elementary volume.}
In the experimental study of dispersion of tracer particles in a 3-D porous-media flow,  \cite{souzy2020velocity} identified a transition from a ballistic regime to an intermediate, anomalous regime, and found that the transition to the asymptotic Fickian regime is determined by the minimal velocity.
There are also studies devoted to investigating the transport and mixing processes in the presence of additional, complex effects, such as the effect of flow inertia \citep{nissan2018inertial} among others. 

{\color{black}The flow pattern and flux in buoyancy-driven porous-media flow have received much attention, for its relevance to various processes in nature and industry, such as geothermal energy recovery and geological sequestration of carbon dioxide \citep{huppert2014fluid,hewitt2020vigorous}.
Recently, it has been found that hydrodynamic dispersion has a significant effect on the flow properties of porous-media convection \citep{hidalgo2009effect,emami2015co2,wen2018rayleigh,de2020non}.
In the related numerical studies, the Fickian dispersion model is commonly used \citep{bear1972dynamics}, which may only be valid at asymptotically large scales.
It is interesting to study when and how the non-Fickian dispersion affects the macroscopic properties of porous-media convection.
The construction and evaluation of macroscopic transport models require a good understanding of the pore-scale transport process.
Thus, it is important to investigate the pore-scale transport behaviour, which very often exhibits anomalous, non-Fickian features.}
The fact that hitherto few studies exist for particle transport in the porous-media convection provides a motivation for the present work.
Here we report a numerical study on the transport of fluid particles and heat transfer in random porous media based on pore-scale modelling. 
In pore-scale models, the detailed flow features in the pores are resolved, which are useful for constructing appropriate macroscopic models and understanding the microscopic mechanisms underlying the macroscopic flow properties \citep{wood2020modeling,gasow2020effects}.

The paper is organized as follows. In \S\ref{sec:numerical}, the model and numerical approaches are described. The main results are presented in \S\S\ref{sec:velocity_statistics}-\ref{sec:heat}, focusing on the flow field, fluid particle transport and heat transfer properties. Finally, summaries of this study are given in \S\ref{sec:summary}.

\section{Numerical model}\label{sec:numerical}
We consider two-dimensional Rayleigh--B{\'e}nard (RB) convection in a square cell.
\makered{The bottom and top plates are heated and cooled, respectively, with a temperature difference $\Delta$.
A simple, model porous medium is included in the cell, which consists of randomly distributed, circular obstacles.}
The fluid flow in the pores is governed by the Oberbeck-Boussinesq equations:
\begin{eqnarray}
\begin{split}
&\frac{\partial T}{\partial t}+\bm{\nabla} \cdot (\bm{v} T)=\frac{1}{\sqrt{Pr R a_f}} \nabla^{2} T, ~~~ \bm{\nabla} \cdot \bm{v}=0, \\
&\frac{\partial \bm{v}}{\partial t}+\bm{v} \cdot \bm{\nabla} \bm{v}+\bm{\nabla} p=\sqrt{\frac{P r}{R a_f}} \nabla^{2} \bm{v}+T \bm{e}_{z}\makered{+\bm{f}}, 
\end{split}
\label{governing_equations}
\end{eqnarray}
where $\bm{v}=(u,w)$ is the velocity vector in the $(x,z)$ plane, $T$ is the temperature and $p$ the pressure. The unit vector $\bm{e}_{z}$ denotes the direction of the buoyancy force.
\makered{The vector $\bm{f}$ in the momentum equation denotes the immersed boundary force to account for the presence of obstacles.}
The dimensionless control parameters are the {\color{black}fluid Rayleigh number} $Ra_f=g \beta \Delta L^{3}/(\nu\kappa)$ and the Prandtl number $~Pr=\nu/\kappa$, where $g$, $\beta$, $\nu$ and $\kappa$ denote the gravitational acceleration, thermal expansion coefficient, kinematic viscosity and thermal diffusivity, respectively.
The cell height $L$, temperature difference $\Delta$ and free-fall velocity $U=\sqrt{g \beta \Delta L}$ are used to non-dimensionalize the governing equations.
Yet another dimensionless parameter is the porosity $\phi$, quantifying the volume fraction of the fluid phase. In the traditional RB convection without porous structure, we have $\phi=1$.
{\color{black} Besides porosity, an additional key non-dimensional number for the porous medium is the Darcy number, $Da=K/L^2$, where $K$ is the permeability, measuring the ability for the fluid to flow through the medium. For porous-media convection, the appropriate Rayleigh number would be the Darcy Rayleigh number $Ra_D=DaRa_f$. For a fixed $Ra_f$, $Ra_D$ is dependent on the medium properties.} The heat transfer efficiency of the system is measured by the Nusselt number, \makered{$Nu=-\langle\partial_z T\rangle_{W,t}$}, where $\langle\cdot\rangle_{W,t}$ denotes taking average over the horizontal wall and over time.

We impose no-slip and no-penetration boundary conditions on the cell boundaries and fluid--obstacle interfaces. For the thermal boundary conditions, the horizontal top and bottom plates are kept at fixed temperatures, and the sidewalls are thermally insulated. The fluid and obstacles are assumed to have the same thermal properties.

The simulation is based on a second-order finite-difference method \citep{verzicco1996finite,van2015pencil}.
\makered{The simulation domain is discretized using a uniform, staggered, Cartesian grid. The time stepping of the explicit terms is based on a fractional-step third-order Runge-Kutta scheme, and the implicit terms based on a Crank-Nicolson scheme.
We employ the direct-forcing immersed boundary approach to account for the obstacles \citep{uhlmann2005immersed,breugem2012second}. 
The moving-least-squares approach is used for the interpolation and spreading procedures between the Eulerian grid and Lagrangian grid \citep{vanella2009short,de2016moving,spandan2017parallel,spandan2018a}.
The heat transfer between the fluid and the obstacles is considered by solving the temperature equations in both phases  \citep{ardekani2018heat,ardekani2018numerical,sardina2018buoyancy}. To do so, a phase indicator $\xi$ is defined based on a level-set function to quantify the solid volume fraction.
Using $\xi$, one can define the velocity $\bm{u}_{cp}$ of the combined phase as
\begin{equation}
\bm{u}_{cp}=(1-\xi)\bm{u}_f+\xi\bm{u}_p,
\end{equation}
with $\bm{u}_f$ and $\bm{u}_p=0$ denoting the velocities in the fluid phase and the obstacles, respectively.}
We refer the reader to \cite{liu2020rayleigh} for more details of the numerical approaches.
Lagrangian tracking algorithm is employed in the direct numerical simulation.
In total, 200$\,$000 fluid particles are tracked. 

{\color{black} For the simulations, a uniform Eularian grid is used with sufficient resolution to resolve the boundary layers and bulk flows, satisfying the classical criterion for the direct numerical simulations of turbulent convection \citep{shishkina2010boundary}. A grid of 1080$\times$1080 is used for most cases. For the smallest porosity with the smallest characteristic pore scale, a grid of 2160$\times$2160 is employed. The circular grains of porous medium are resolved with at least 22 grid nodes.}

In this study, the transport properties of fluid particles and heat transfer are investigated for {\color{black}$\phi\in[0.654,1]$} \makered{in order to study the transition of flow behaviours from the classical turbulent Rayleigh--B{\'e}nard state.}
\makered{We consider fluid Rayleigh number $Ra_f=10^8,~10^9$ and the Prandtl number $Pr$ at 4.3, which is a typical value for water at $40^{\text{o}}\text{C}$.}
\makered{At the Rayleigh numbers considered, the driving buoyancy force is large enough to overcome the enhanced friction due to the porous structure for the range of porosity we considered, and the resulting temporal and spatial fluctuations of flow field could play an important role on the particle transport.}

\makered{A set of random porous media with varying porosities is constructed by gradually increasing the number of obstacles in the cell. The obstacles are not overlapping with the pore scale no less than a minimum value $l_{min}=0.005$ for most configurations. For the one with the smallest porosity $\phi=0.654$, $l_{min}=0.004$ is used to accommodate more obstacles in the cell. 
For simplicity, we consider only monodisperse obstacles with a fixed diameter $D=0.02$. 
In order to examine the influence of pore layout on the flow behaviours, two different sets of random porous media are considered. Similar flow patterns and statistics are obtained based on the two sets of porous media, demonstrating the robustness of the flow behaviours with respect to the details of obstacle arrangement. In the following sections, we just show the results for one set of layout.}
A reasonable estimate of $Da$ can be obtained based on the Kozeny's equation $Da=\phi^3D^2/[\beta(1-\phi)^2L^2]$ \citep{nield2006convection,gasow2020effects}. With the empirical model coefficient $\beta=150$, the minimum Darcy number reached in this study is $Da=6.2\times10^{-6}$.
\makered{The simulations were run over at least 1000 time units after the initial transients to obtain good statistical convergence. The relative difference $e_{Nu}$ of $Nu$ based on the first and second halves of the simulations is generally less than 1\%, except in the case with the smallest porosity $\phi=0.654$ where the simulation is more computationally demanding and corresponding $e_{Nu}$ is approximately 3\%.}

\section{Flow field}\label{sec:velocity_statistics}
\begin{figure}
	\centering
	\includegraphics[width=\linewidth]{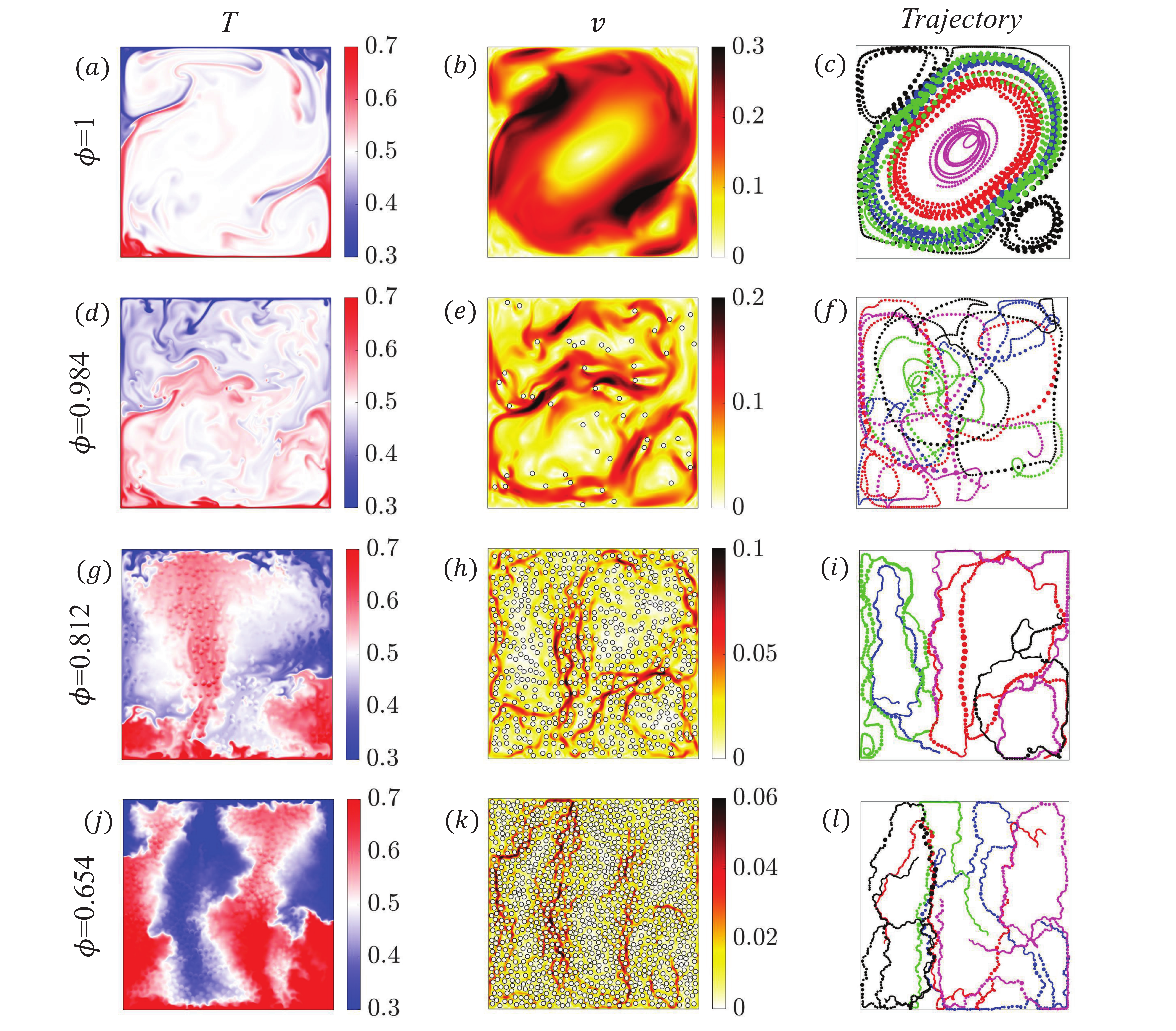}
	\vspace{-3 mm}
	\caption{\label{flow_structures}Snapshots of ($a,d,g,j$) temperature $T$ and ($b,e,h,k$) velocity magnitude $v\equiv|\bm{v}|$, and $(c,f,i,l)$ the typical trajectories of fluid particles for various $\phi$ at the same fluid Rayleigh number $Ra_f=10^9$: $(a-c)$ $\phi=1$, ($d-f$) $\phi=0.984$, $(g-i)$ $\phi=0.812$ and $(j-l)$ $\phi=0.654$. In the snapshots of $v$, the obstacles are indicated by circles. In panels $(c,f,i,l)$, the trajectories of five fluid particles are depicted, indicated by different colours. Particle positions are shown with fixed time intervals in each plot. The magnitude of particle velocity is quantified by the marker size, with larger marker for larger velocity. \makered{The integration times for the particle trajectories in panels ($c,f,i,l$) are 50, 80, 150 and 400 free-fall time units, respectively.}}
\end{figure}

Figure \ref{flow_structures} shows the instantaneous fields of the temperature $T$ and velocity magnitude $v\equiv|\bm{v}|$, and the typical trajectories of fluid particles at various $\phi$.
In the traditional RB convection with $\phi=1$, the flow consists of a well-organized large-scale circulation and two corner rolls, around which the fluid particles take a quite periodic motion, as depicted in figure \ref{flow_structures}$(c)$.
We find that, in  the presence of obstacles, the flow organization and particle movement are strongly modified. With the decrease of $\phi$, the convection strength is reduced due to the enhanced drag of the porous matrix, and the temperature mixing in the bulk is less efficient, as revealed by the snapshots of the temperature and velocity magnitude in figure \ref{flow_structures}.
In the presence of a small number of obstacles, the flow structure is less organized, and consequently the fluid trajectories become more irregular, as shown in figure \ref{flow_structures}$(f)$.
The flow field in porous media is highly heterogeneous due to the interaction between the porous medium and coherent flow structures, and it fluctuates both spatially and temporally. Interestingly, for small $\phi$, the flow is dominated by large-scale plumes, and convection channels with strong flow strength emerge in the pores, which are closely related to the plume dynamics, and the patches with low velocities appear due to the impedance of the obstacles (see figures \ref{flow_structures}$(h,k)$). Along the convection channels, long-range transport of fluid particles is observed, while the particles can stay `trapped' for relatively long duration in the low-velocity regions.
{\color{black}For small $\phi$, the flow exhibits columnar structures, similar to those observed in the severely confined RB convection \citep{chong2016exploring}. }

\begin{figure}
	\centering
	\vspace{0 mm}
	\includegraphics[width=0.85\linewidth]{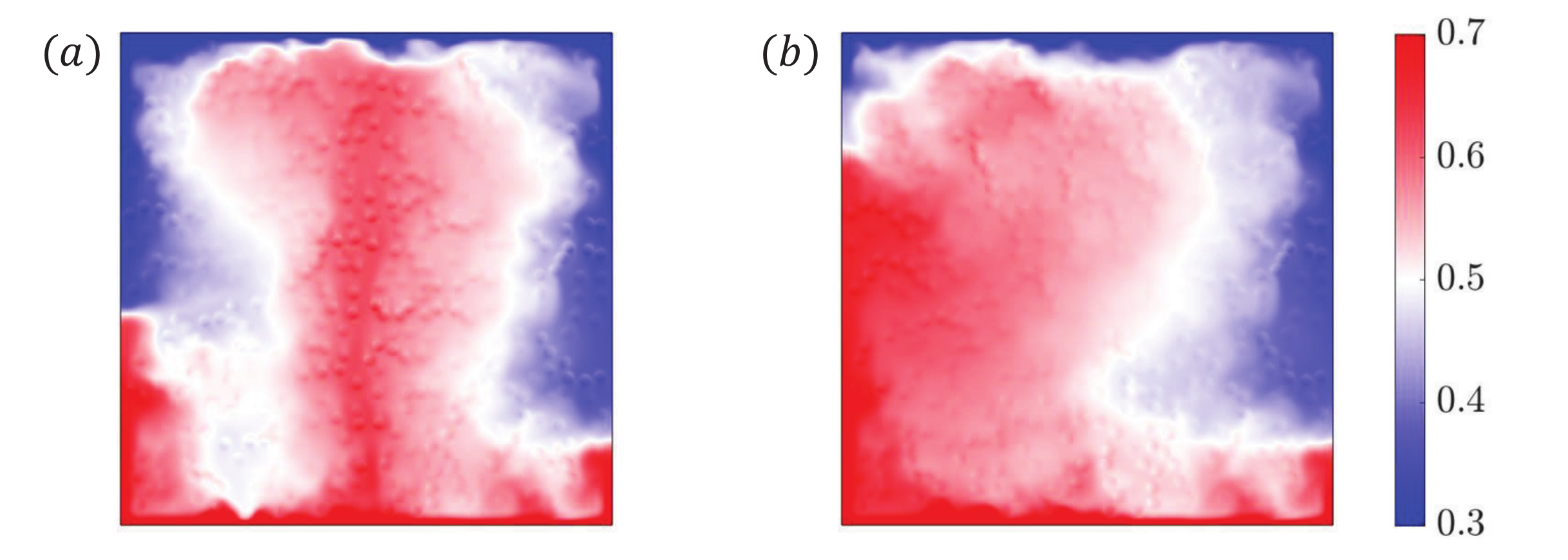}
	\caption{\label{multiple_states}{\color{black}Time-averaged temperature fields} at $Ra_f=10^8$ and $\phi=0.859$ obtained with two different initial conditions. The two solutions are persistent over at least 1000 time units, {\color{black}and the mean fields are obtained by averaging more than 5000 snapshots.} The Nusselt numbers corresponding to panels $(a,b)$ are 25.4 and 24.1, respectively.}
\end{figure}

\makered{ We note that the depicted flow patterns in figure \ref{flow_structures} are robust over the time of present simulations, in the sense that, although the flow details in the pores vary with time, the global structure of flow pattern is persistent. 
However, in a porous medium, the statistical steady states may be different for different initial conditions. Due to the suppression of fluctuations in the presence of obstacles, the solution may not be able to sample all the flow configurations over the time range considered.
As an example, we show in figure \ref{multiple_states} two {\color{black}time-averaged temperature fields for the same parameters of $Ra_f=10^8$ and $\phi=0.859$}, which are obtained with two different initial conditions.
Despite the existence of multiple states, we expect that the differences of the statistics of multiple flow states are relatively small, and the global trend of variation of flow behaviours with porosity will not be qualitatively affected by the appearance of multiple states.}

\begin{figure}
	\centering
	\includegraphics[width=0.48\linewidth]{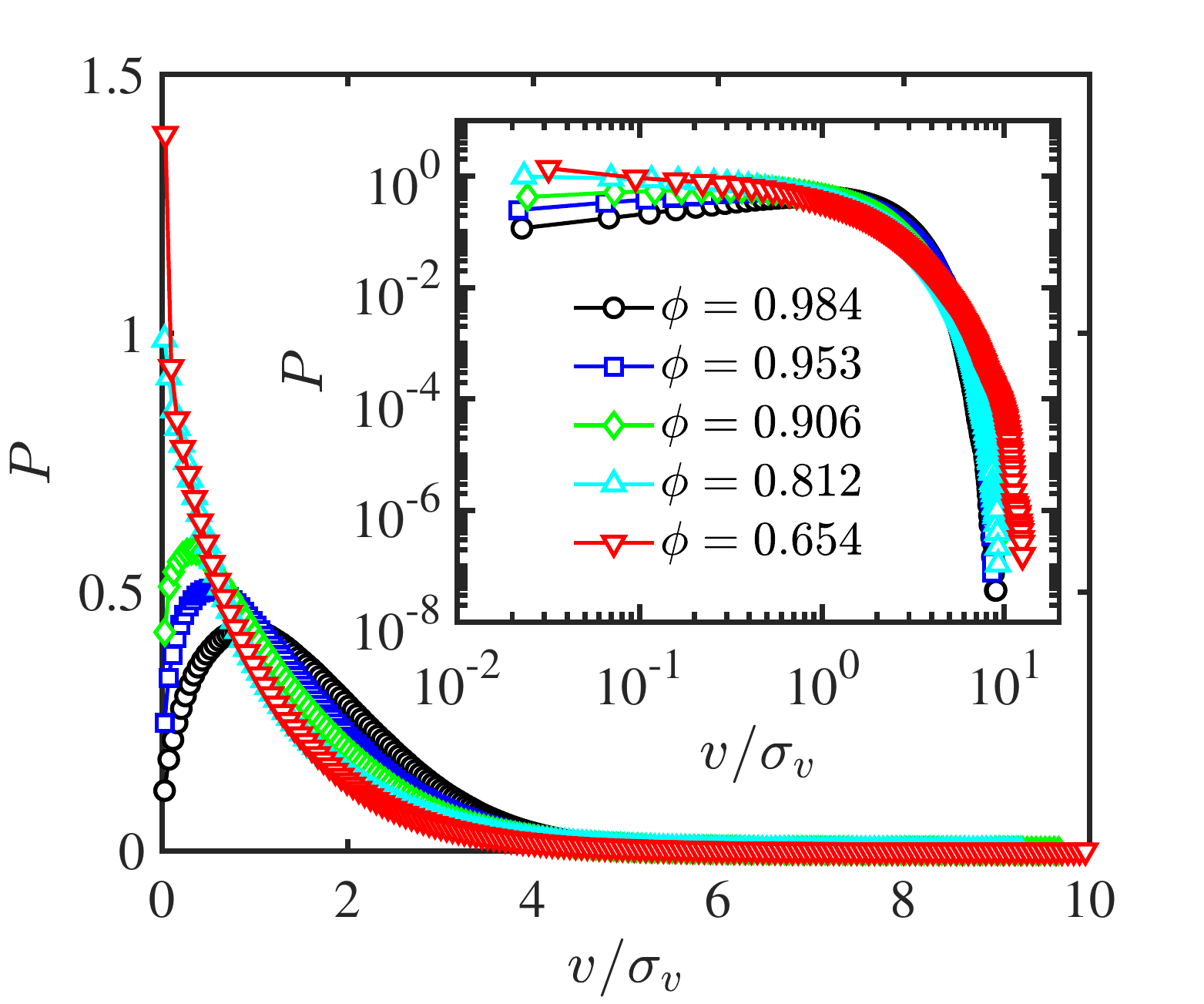}
	\put(-185,136){$(a)$}
	\hspace{0 mm}
	\includegraphics[width=0.48\linewidth]{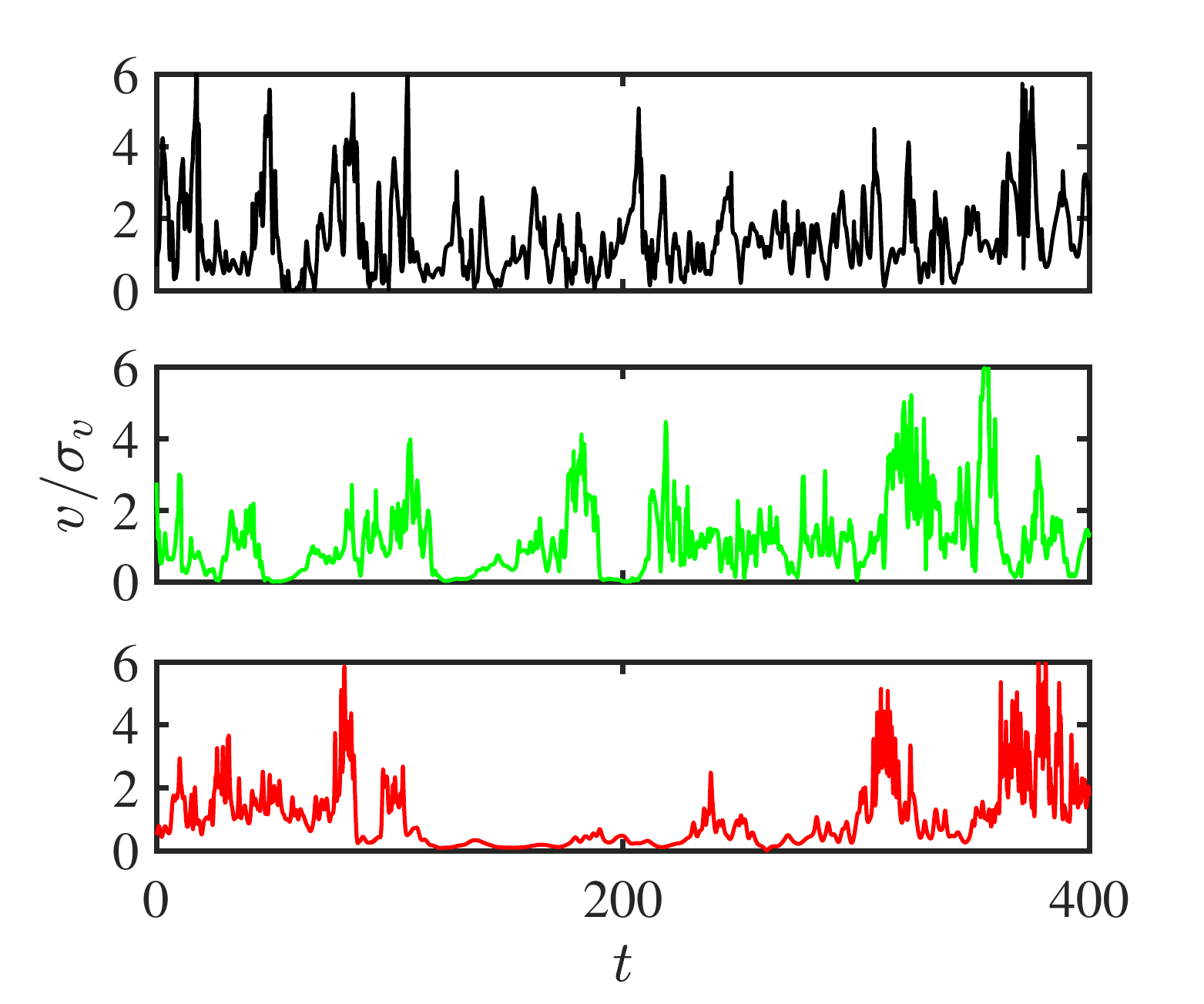}
	\put(-183,136){$(b)$}
	\vspace{0 mm}
	\caption{\label{velo_pdf_correlation}$(a)$ The p.d.f.s $P(v/\sigma_v)$ of the normalized \makered{Lagrangian} velocity magnitude $v/\sigma_v$ for various $\phi$, where $\sigma_v$ denotes the standard deviation of $v$. The inset shows \makered{the same data in a double logarithmic scale. $(b)$ Typical time series of $v/\sigma_v$ for three values of $\phi$, which are indicated by the colours. From top to bottom, the porosities are 0.984, 0.906 and 0.654, respectively.}}
\end{figure}

The particle movement is highly irregular in the heterogeneous flow field of thermal convection in porous media. In order to quantify the chaotic particle dynamics, we plot the p.d.f.s $P(v/\sigma_v)$ of the normalized \makered{Lagrangian} velocity magnitude $v/\sigma_v$ for various $\phi$ in \makered{figure \ref{velo_pdf_correlation}$(a)$}, where $\sigma_v$ is the standard deviation of $v$. As $\phi$ is decreased, the shape of $P(v/\sigma_v)$ is significantly modified. 
With the decrease of $\phi$, the probability density for low velocity increases significantly.
The time series of the normalized velocity magnitude $v/\sigma_v$ of a fluid particle are shown in \makered{figure \ref{velo_pdf_correlation}$(b)$}.
We find that for small $\phi$, due to the emergence of the convection channels and low-velocity regions, the time variation of $v$ exhibits strong intermittency, with the existence of high-velocity bursts, interrupted by long-time trapping events with low velocities.
\makered{Similar variations of Lagrangian velocity are observed based on the other set of porous-media layout, demonstrating the robustness of the statistics with respect to the details of pore layout.}

\begin{figure}
	\centering
	\includegraphics[width=0.6\linewidth]{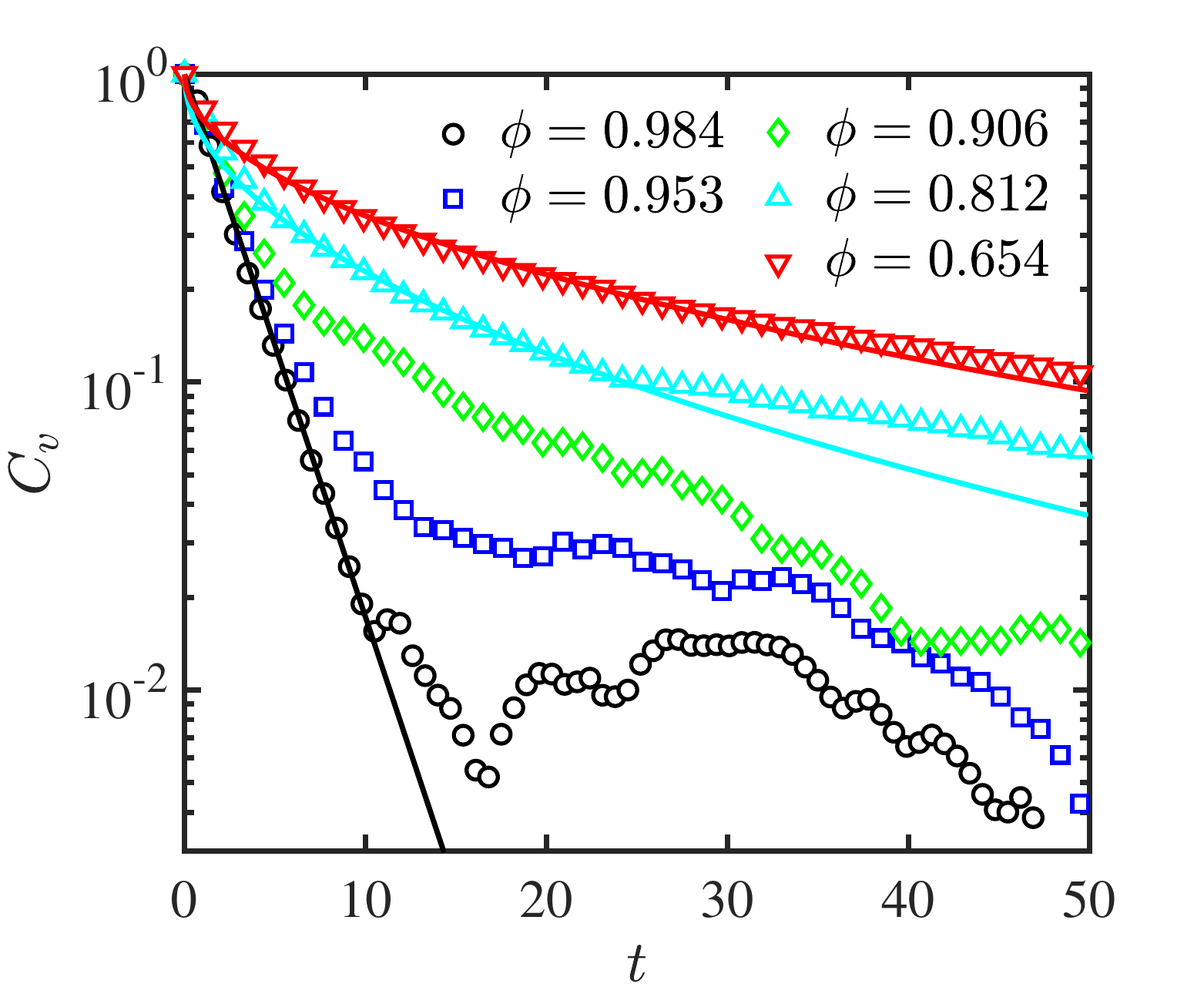}
	\vspace{-1 mm}
	\caption{\label{correlation}\makered{Temporal autocorrelation function $C_v$ of Lagrangian velocity magnitude $v$ for various $\phi$. The three lines denote the fittings of the initial parts of $C_v(t)$ by an exponential function $e^{-at} $ (black line) and stretched-exponential function $e^{-\sqrt{bt}}$ (cyan and red lines), where $a$ and $b$ are the fitting parameters.}}
\end{figure}

We plot in \makered{figure \ref{correlation}} the temporal autocorrelation function $C_v(t)=\langle [v_0-\langle v_0\rangle][v_t-\langle v_t\rangle]\rangle/\sigma_{v}^2$ of the particle velocity $v$ to further quantify the particle dynamics, {\color{black}where $v_0$ and $v_t$ indicate the velocity magnitudes at times 0 and $t$, respectively. The symbol $\langle\cdot\rangle$ \makered{denotes averaging} over an ensemble of fluid particles.}
\makered{Results based on the two sets of random porous-media layouts are consistent.}
\makered{Figure \ref{correlation} shows that} $C_v$ decreases with $t$, demonstrating the loss of memory of the particle dynamics.
Compared with the large-$\phi$ case, $C_v$ decays slower for smaller $\phi$, and the autocorrelation is enhanced with decreasing $\phi$ at fixed $t$.
We note that, as $\phi$ is decreased, the decay form of $C_v$ is qualitatively changed, besides the quantitative differences in the decaying rate.
For large $\phi$, the initial part of the decay process takes an exponential form, as demonstrated by the linear behaviour, $ln(C_v)\sim t$, in the log-linear plot. For the two smallest $\phi$, the decaying behaviours of $C_v(t)$ are similar and approximately follow a stretched-exponential form with $ln(C_v)\sim \sqrt{t}$, which suggests that the change of the velocity autocorrelation at small enough $\phi$ is attributed to the change of the characteristic time scale of Lagrangian fluid transport.

The relaxation behaviour in a stretched-exponential form at small $\phi$ is attributed to the existence of very large relaxation times. To show this, the stretched exponential may be expressed as an integral of exponential functions with a spectrum of decaying rates \citep{bouchaud2008anomalous,johnston2006stretched}:
\begin{equation}
\centering
\exp\left[-\left(\frac{t}{\tau}\right)^{1/2}\right]=\int_{0}^{\infty}P(\lambda)\exp\left(-\lambda\frac{t}{\tau}\right)d\lambda,
\label{sum_of_exponential_functions}
\end{equation}
where $\tau$ denotes the characteristic relaxation time, $\lambda$ is the ratio of relaxation time and $P(\lambda)$ is the p.d.f. of $\lambda$. Smaller $\lambda$ indicates slower relaxation. Since the right-hand side of (\ref{sum_of_exponential_functions}) is the Laplace transform of $P(\lambda)$, $P(\lambda)$ is obtained via taking the inverse Laplace transform. For small $\lambda$, one obtains \citep{johnston2006stretched}
\begin{equation}
\centering
P(\lambda)\approx \exp\left(-B\lambda^{\beta}\right)
\label{pdf_relaxation_time}
\end{equation}
up to subleading power-law corrections, where $\beta=-1$, and $B$ is a positive constant. Expression (\ref{pdf_relaxation_time}) suggests that a very limited probability of super-slow relaxation can result in a global relaxation in a stretched-exponential form \citep{bouchaud2008anomalous}. The appearance of large relaxation times is attributed to the trapping events of fluid particles at low-velocity regions. The purely exponential decay corresponds to $\beta=-\infty$, i.e., no long-time relaxation \citep{bouchaud2008anomalous}.

\section{Fluid particle transport}\label{sec:transport}
\begin{figure}
	\centering
	\includegraphics[width=0.6\linewidth]{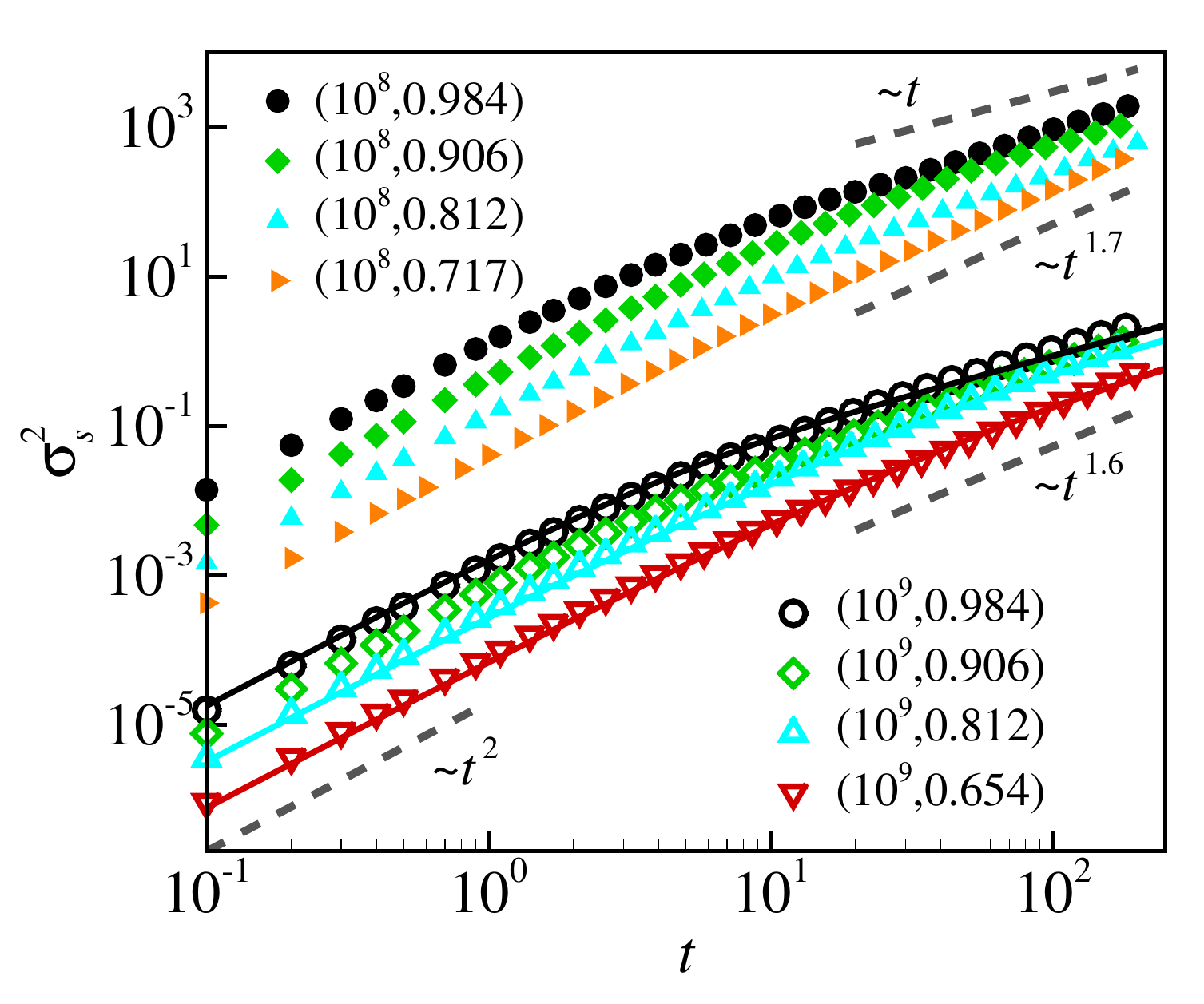}
	\vspace{-1 mm}
	\caption{\label{displacement_variance} Displacement variance $\sigma^2_{s}(t)$ of fluid particles along the trajectory for various $\phi$ at $Ra_f=10^8$ and $10^9$. The values of parameters $(Ra_f,\phi)$ are given in the legend. The results at $Ra_f=10^8$ are shifted upward for clarity. The solid lines are the growth curves of $\sigma_s^2(t)$ obtained from (\ref{variance_correlation}) and the idealized velocity autocorrelation functions in exponential (black) and stretched-exponential (cyan and red) forms. For reference, several scaling laws are included as dashed grey lines.}
\end{figure}

Now we study the transport properties of fluid particles by quantifying the particle displacement $s(t)\equiv\int_{0}^{t}v(\tau)d\tau$ along the trajectory.
The transport behaviours for the ensemble of particles exhibit strong fluctuations, and the ensemble-averaged displacement $\langle s\rangle$ grows with the time-independent mean velocity $\langle v\rangle$, namely, $\langle s\rangle=\langle v\rangle\cdot t$.
\makered{The width of the displacement distribution is measured by the displacement variance $\sigma^2_{s}=\langle (s-\langle s\rangle)^2\rangle$.}
Figure \ref{displacement_variance} shows the time evolution of $\sigma^2_{s}$ of fluid particles for different $\phi$ at $Ra_f=10^8$ and $10^9$.  It is found that $\sigma^2_{s}$ grows ballistically with $\sigma^2_s\sim t^2$ at small $t$. This ballistic regime is robust for different $\phi$, while the proportionality constant decreases with $\phi$.
The ballistic regime is a universal phenomenon for general transport processes \citep{batchelor1950application,bourgoin2015turbulent,mathai2018dispersion}, and is associated with the strong Lagrangian velocity autocorrelations observed at small times.

Figure \ref{displacement_variance} also shows the deviation from the ballistic regime at large $t$, and $\sigma_s^2$ exhibits a sub-ballistic scaling, $\sigma^2_s\sim t^{\gamma}$, with an effective scaling exponent $\gamma<2$. 
The transition to a different transport behaviour at large $t$ is expected and indicates the loss of memory of fluid particles to the initial conditions \citep{bourgoin2015turbulent}.
In the presence of a small number of obstacles, {\color{black} due to the impact of strong velocity fluctuations,} the fluid particles can efficiently explore the irregular flow field and lose the memory about the initial conditions. Consequently, the ballistic regime terminates at an early time, and beyond that the fluid particles approximately exhibit a Fickian transport behaviour with $\sigma^2_s\sim t$, as in the case of $\phi=0.984$ in figure \ref{displacement_variance}.
When $\phi$ is further decreased, $\sigma_s^2$ at relatively large time displays a clear deviation from the Fickian behaviour, with an effective scaling exponent $1<\gamma<2$.
This anomalous non-Fickian behaviour of particles in porous media is associated with the increased probability density of low velocity \citep{berkowitz2006modeling,souzy2020velocity}.
Similar phenomena are observed for both values of $Ra_f$ considered, demonstrating the robustness for the emergence of non-Fickian behaviour with the decrease of $\phi$ in porous-media convection.
Non-Fickian behaviour of particle transport is an omnipresent phenomenon and has been reported in many different settings, both with and without porous media \citep{richardson1926atmospheric,grossmann1990diffusion,berkowitz2006modeling,salazar2009two,bijeljic2011signature,bourgoin2015turbulent,puyguiraud2019stochastic,puyguiraud2019upscaling,dentz2018mechanisms,souzy2020velocity,taghizadeh2020preasymptotic}.

The transport properties of fluid particles can be related to the temporal autocorrelation function $C_v$ of particle velocity $v$, in the spirit of the Green-Kubo relations, which connect a transport coefficient to a correlation function in time \citep{kubo2012statistical}.
Considering that $\langle s\rangle=\langle v\rangle\cdot t$, $\sigma_s^2$ can be directly related to $C_v$ as
\begin{eqnarray}
\begin{split}
\sigma_s^2(t)&=\langle (s-\langle s\rangle)^2\rangle=\langle (\int_{0}^{t}dt'[v(t')-\langle v\rangle])^2\rangle \\
&=2\sigma_v^2\int_{0}^{t}dt'(t-t')C_v(t').
\end{split}
\label{variance_correlation}
\end{eqnarray}
Some derivation details are given in Appendix \ref{appendix1}.
Based on the numerically identified correlation behaviours in \makered{figure \ref{correlation}}, here we consider two empirical, idealized forms of the autocorrelation function: $C_{v,1}(t)=e^{-at}$ and $C_{v,2}(t)=e^{-\sqrt{bt}}$, where $a$ and $b$ are the corresponding decaying rates. Here $C_{v,1}$ and $C_{v,2}$ are in the exponential and stretched-exponential forms, which are similar in form to the correlation properties of fluid particles in porous media with large and small $\phi$, respectively. 
Based on (\ref{variance_correlation}) and symbolic integration we obtain
\begin{equation}
\left\{
\begin{aligned}
&\sigma^2_{s,1}(t)=\frac{2\sigma^2_v}{a^2}(at-1+e^{-at}),\\
&\sigma^2_{s,2}(t)=\frac{4\sigma^2_v}{b^2}[bt-6+2e^{-\sqrt{bt}}(bt+3\sqrt{bt}+3)].
\end{aligned}
\right.
\label{GK_porous}
\end{equation}
Despite that the assumed autocorrelation functions are highly idealized, the global trend of variation of $\sigma_s^2(t)$ can be captured by (\ref{GK_porous}), as shown in figure \ref{displacement_variance}. 
Expressions (\ref{GK_porous}) show that, for both forms of the velocity autocorrelation function, the particles will reach the Fickian regime in the long-time limit. The appearance of the initial ballistic regime is evident from the series expansions of (\ref{GK_porous}) at $t=0$.
When the particle velocities are exponentially correlated in time, the particles will exhibit Fickian transport behaviour for $t\gg 1/a\sim O(1)$;
while when the autocorrelation function $C_v$ has a perfect, stretched-exponential form, $C_v=e^{-\sqrt{bt}}$, the term proportional to $e^{-\sqrt{bt}}$ may disrupt the appearance of the Fickian behaviour even when $t\gg1$, resulting in the anomalous transport behaviour at relatively large $t$, consistent with the observation in figure \ref{displacement_variance}. 
Deviations of (\ref{GK_porous}) from the simulation results are visible, which are attributed to the deviations of the velocity autocorrelation of fluid particles from the idealized exponential or stretched-exponential forms, as shown in \makered{figure \ref{correlation}}. 
Since the idealized autocorrelation functions underestimate the autocorrelation of particle velocity in porous media at large $t$ for the parameters \makered{shown}, the predicted displacement variance (\ref{GK_porous}) may fail to capture the behaviour of $\sigma_s^2$ for large times.

The above analysis shows that the anomalous transport behaviour at small $\phi$ is associated with the qualitatively different time correlation properties of fluid particles.
When $t$ is large enough, the term proportional to $t$ will dominate over other terms, and the Fickian transport behaviour will be achieved \citep{souzy2020velocity}.
Considering that the decaying rate of $C_v(t)$ is expected to decrease for smaller $\phi$, the preasymptotic, anomalous transport behaviour may last for relatively long time compared with the characteristic time scale of particle transport at small $\phi$, and is expected to be relevant in realistic porous-media convection.

\section{Heat transfer}\label{sec:heat}

\begin{figure}
	\centering
	\includegraphics[width=0.46\linewidth]{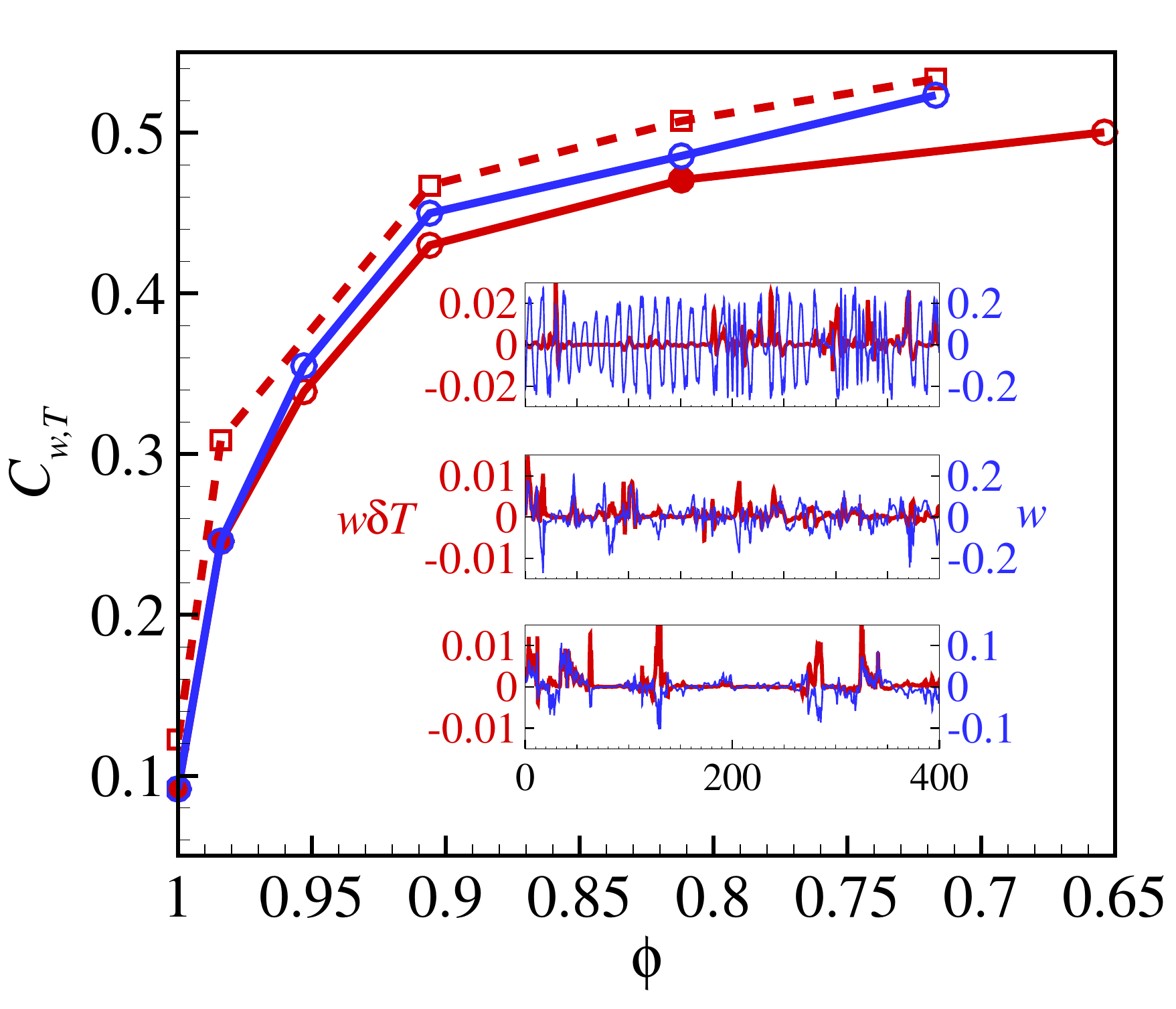}
	\put(-176,136){$(a)$}
	\hspace{4 mm}
	\includegraphics[width=0.46\linewidth]{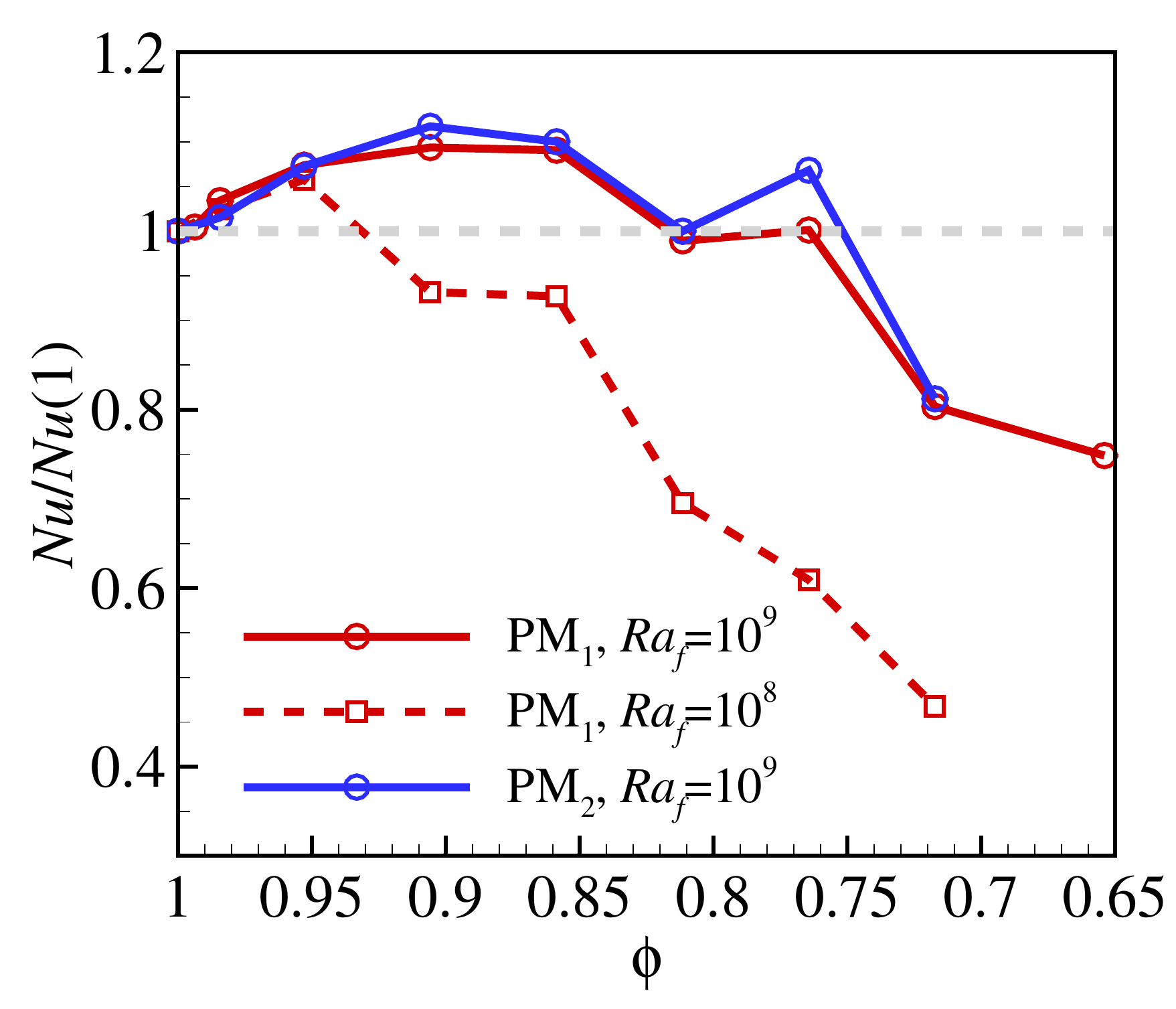}
	\put(-176,136){$(b)$}
	\vspace{-1 mm}
	\caption{\label{heat_statistics} \makered{Heat transfer statistics based on two different sets of random porous-media layouts $\text{PM}_1$ and $\text{PM}_2$.} $(a)$ Cross-correlation $C_{w,T}$ between the vertical velocity $w$ and temperature $T$ as a function of $\phi$ for fluid particles in the the whole cell at $Ra_f=10^8$ and $10^9$. The inset shows the typical time series of the convective heat flux $w\cdot \delta T$ in the vertical direction (red lines) and $w$ (blue lines) \makered{based on $\text{PM}_1$} at $Ra_f=10^9$ and various $\phi$: from top to bottom $\phi=1$, $0.984$ and $0.812$, corresponding to the filled symbols in the $C_{w,T}(\phi)$ plot. $(b)$ \makered{The normalized Nusselt number $Nu/Nu(1)$} as a function of $\phi$ for $Ra_f=10^8$ and $10^9$.}
\end{figure}

Now we focus on the heat transfer properties of the fluid particles.
We plot in \makered{figure \ref{heat_statistics}$(a)$} the cross-correlation $C_{w,T}$ between the vertical velocity $w$ and temperature $T$ as a function of $\phi$ for fluid particles in the whole cell. We find that $C_{w,T}$ becomes significantly larger with the decrease of $\phi$.
The small value of $C_{w, T}$ in the traditional RB convection is attributed to the uniform temperature distribution in the bulk with $T\approx T_m$, where $T_m$ is the arithmetic mean temperature.
The large value of $C_{w,T}$ at small $\phi$ indicates that the velocity field is closely related to the motion of thermal plumes in porous-media convection, and the interaction between the porous media and thermal plumes results in the heterogeneous velocity fields.
The increase of $C_{w,T}$ also reveals a remarkable mechanism to enhance the overall heat transfer of thermal convection with porous structure.
{\color{black}As $\phi$ is decreased from 1, heat transfer enhancement is indeed observed, as shown in \makered{figure \ref{heat_statistics}$(b)$}. When $\phi$ is small enough, heat transfer is reduced compared with that of traditional RB convection, which is attributed to the strong suppression of convection due to the additional drag of porous medium, as discussed in \citep{liu2020rayleigh}.}
\makered{The anomalous transition of $Nu$ with $\phi$ around $\phi=0.764$ is associated with the change of flow organization.}

The inset of \makered{figure \ref{heat_statistics}$(a)$} shows the time series of the convective heat flux $w \delta T$ in the vertical direction and the vertical velocity $w$ of fluid particles for various $\phi$, where $\delta T=T-T_m$.
In the traditional RB convection with $\phi=1$, the fluid particles with high velocity may not contribute to the vertical heat transfer for relatively long times, which is due to the fact that the temperature in the centre core is well mixed with small temperature fluctuation $\delta T$; while for small $\phi$, the low amplitude of the vertical heat transfer is mainly due to the low particle velocity, confirming the strong cross-correlation between the vertical velocity and temperature fluctuation. As a consequence of this enhanced cross-correlation, the probability of negative Lagrangian heat transfer in the bulk region will be decreased compared with that in the traditional RB convection.

\begin{figure}
	\centering
	\includegraphics[width=0.85\linewidth]{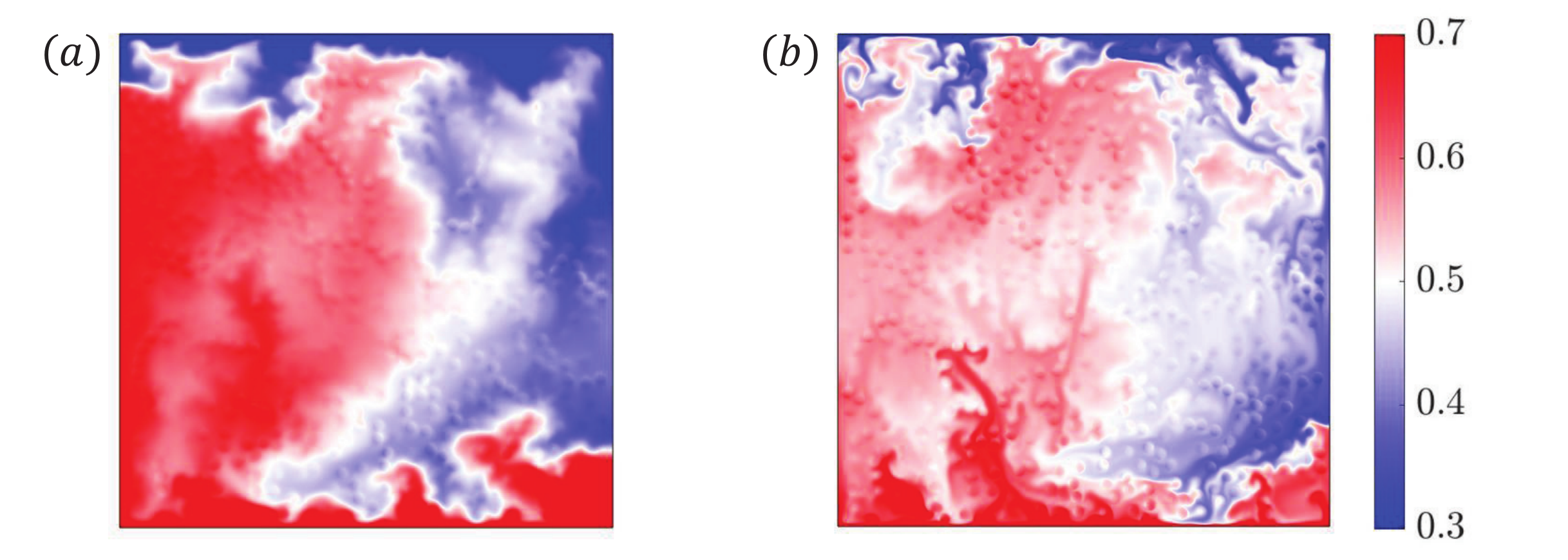}
	\caption{\label{peclet}\makered{Instantaneous temperature distributions at $(a)$ $Ra_f=10^8$ and $(b)$ $Ra_f=10^9$ at $\phi=0.812$.}}
\end{figure}

The pore-scale Peclet number $Pe=vD/\kappa$ quantifies the ratio of the thermal diffusion time scale $\tau_{\kappa}=D^2/\kappa$ to the convection time scale $\tau_c=D/v$, where $v$ is the fluid velocity, $D$ the obstacle diameter and $\kappa$ the thermal diffusivity. In dimensionless form, we have $\kappa=1/\sqrt{PrRa_f}$.
In the convection channels with fast fluid transport, $Pe$ is high, and in the low-velocity regions, $Pe$ is low. For the Rayleigh numbers considered, $Pe$ is generally larger than 1 (except at the stagnant regions with very low velocities). For $Ra_f=10^9$, the peak value of $Pe$ is of order $O(100)$ at small porosities. The Peclet number has important implications to the temperature evolution. A large $Pe>1$ indicates that the transport of temperature is dominated by convection rather than diffusion, such that there may not be enough time for the obstacles to respond to the temperature fluctuations, which may result in a less uniform temperature distribution across the fluid-solid interface. A larger $Pe$ (stronger convection) will also result in more efficient convective mixing.
As an example, we depict in figure \ref{peclet} the instantaneous temperature distributions for $Ra_f=10^8$ and $10^9$ at $\phi=0.812$. It is found that the bulk temperature for larger $Ra_f$ (larger $Pe$) is closer to the arithmetic mean value $T_m=0.5$, demonstrating the more efficient temperature mixing for larger $Ra_f$. When the bulk flow is sufficiently mixed, one would expect a uniform temperature distribution with $T\approx T_m$, as shown in figure \ref{flow_structures}$(a)$. Figure \ref{peclet} also shows that for larger $Ra_f$ the temperature across the fluid--solid interface is less uniform, as manifested by the more evident imprint of the solid phase in figure \ref{peclet}$(b)$.

\section{Conclusion}\label{sec:summary}
\makered{In summary, we have performed a numerical study of Lagrangian dynamics and heat transfer of buoyancy-driven convection in random porous media. We consider simple model porous media, which consist of randomly distributed circular obstacles in a conventional Rayleigh--B{\'e}nard cell, and the results are obtained for fixed values of obstacle diameter $D$ and specified minimum pore size $l_{min}$.}
We find that {\color{black}the flow field in porous media is highly heterogeneous, with the presence of convection channels with strong fluid transport and stagnant regions with low velocities.} \makered{Similar behaviours have also been reported in other porous-media systems. The emergence of low velocities and stagnation zones are at the origin of the strong intermittency of the Lagrangian velocity series.}
The displacement variance $\sigma_s^2$ of fluid particles shows the existence of distinct transport regimes. For small $t$, a ballistic regime is identified, which transitions to a sub-ballistic regime at large $t$. For small $\phi$, we observe that the fluid particle is able to exhibit anomalous transport with a super-linear growth of $\sigma_s^2$ at relatively long times.
The anomalous transport is associated with the long-time correlation of Lagrangian velocity of the fluid in porous media with small $\phi$. Even though the anomaly is expected to cross over into the Fickian regime in the long-time limit, it may last for a significantly long time when compared with the characteristic time scale of particle transport, and can be of serious consequences in realistic porous-media convection.
Regarding the heat transfer properties, the cross-correlation $C_{w,T}$ between the vertical velocity and temperature fluctuation is significantly enhanced with decreasing porosity, implying the close relation between the plume dynamics and heterogeneous velocity field in porous-media convection.

\makered{
Compared with pressure driven flows, one distinctive characteristic of buoyancy-driven convection is that the driving force is governed by the temperature field, resulting in a strong correlation between the velocity field and temperature field. 
Particularly, cold and warm plumes constantly emerge from the thermal boundary layers on the top and bottom plates and move through the pores under the action of strong buoyancy force, resulting in convection channels with fast fluid transport. Thus, besides the influences of porous structures \citep{alim2017local,de2017prediction}, the dynamics of plumes also plays an important role on the flow pattern and statistics.}

\makered{The present study based on pore-scale modelling sheds some light on the particle transport in buoyancy-driven porous-media convection and highlights the active role of temperature field and thermal plumes to the flow behaviours.}
The results on heat transfer suggest a new approach for enhancing heat transport by controlling the coherence of the bulk flow.
{\color{black}In the future, it is interesting to extend the work to larger parameter space and to more complex porous media, such as those with broad distributions of pore scales \citep{bijeljic2013predictions,gjetvaj2015dual}.}
\makered{Another critical question for future studies is to formulate macroscopic models of porous-media convection with the anomalous dispersion effects taken into account \citep{metzler2000random}.}

\section*{Acknowledgments}
We thank D. Lohse, V. Mathai, Y. Yang and Z. Wan for fruitful discussions.
This work was supported by the Natural Science Foundation of China (Grant Nos. 11988102, 91852202 and 11861131005) and Tsinghua University Initiative Scientific Research Program (Grant No. 20193080058). S.L. acknowledges the project funded by the China Postdoctoral Science Foundation (Grant No. 2019M660614) and Key Laboratory of Advanced Reactor Engineering and Safety of Ministry of Education (Grant No. ARES-2019-10).

\section*{Declaration of interests}
The authors report no conflict of interest.

\appendix

\section{}\label{appendix1}
In this Appendix, some details for the derivation of (\ref{variance_correlation}) are given.
Considering that $\langle s\rangle=\langle v\rangle\cdot t$, the variance $\sigma_s^2(t)$ can be directly related to the autocorrelation function $C_v(t)$ of particle velocity as
\begin{eqnarray}
\begin{split}
&\sigma_s^2(t)=\langle (s-\langle s\rangle)^2\rangle=\langle (\int_{0}^{t}dt'[v(t')-\langle v\rangle])^2\rangle \\
&~~~~~~~=\int_{0}^{t}dt'\int_{0}^{t}dt''\langle [v(t')-\langle v\rangle][v(t'')-\langle v\rangle] \rangle\\
&~~~~~~~=\sigma_v^2\int_{0}^{t}dt'\int_{0}^{t}dt''C_v(t'-t''),
\end{split}
\label{variance_correlation_app}
\end{eqnarray}
where the property of time translation invariance is invoked. By introducing a change of variable, $h=t'-t''$, and employing the symmetry property, $C_v(h)=C_v(-h)$, we obtain
\begin{eqnarray}
\begin{split}
\sigma_s^2(t)=2\sigma_v^2\int_{0}^{t}dh(t-h)C_v(h).
\end{split}
\label{GK_app}
\end{eqnarray}


\end{document}